\documentclass[a4paper]{article}

\usepackage{17lomcon}        
\usepackage{cite}             
\usepackage{graphicx}
\usepackage{pgf}

\begin{document}


\title{LATEST RESULTS FROM T2K}

\author{Marcela Batkiewicz \email{Marcela.Batkiewicz@ifj.edu.pl}, for the T2K collaboration }

\affiliation{Institute of Nuclear Physics Polish Academy of Sciences, Cracow, Poland }

\date{}
\maketitle


\begin{abstract}

The T2K (Tokai to Kamioka) experiment is a long-baseline accelerator neutrino oscillation
experiment. Intense muon neutrino and antineutrino beams are produced at the
J-PARC accelerator complex situated in Tokai. After 280 metres the beam  goes
through the set of two near detectors: INGRID and ND280, where the beam
parameters before oscillations are measured and neutrino cross-sections are
studied. After another 295 km, the distance corresponding to the first oscillation maximum, the neutrino
beam passes through the Super-Kamiokande far detector. Comparison of the beam
characteristics in the near and far detectors makes it possible to
determine the oscillation  parameters.

Based on muon neutrino disappearance, T2K has delivered the world's leading
measurement of the $\theta_{23}$ mixing angle. It has been also the first experiment
to observe electron neutrino appearance, with a significance of 7.3$\sigma$.
These results made it possible to determine the $\theta_{13}$ mixing angle and to
provide the first hint of a non-zero value of the $\delta_{CP}$ phase. The T2K
experiment has also delivered several neutrino cross-section measurements at
neutrino energies around 1 GeV. Currently, T2K is collecting data with a muon
antineutrino beam, for the $\delta_{CP}$ and antineutrino cross-section measurements.

A summary of the most recent oscillation measurements, including the new electron antineutrino
appearance results, is presented.

\end{abstract}


\section{Neutrino oscillations}

Neutrino oscillations is a quantum-mechanical phenomenon in which neutrinos change their flavour from one to another while propagating in space.
This results from the fact that the neutrino flavour eigenstates ($\nu_e, \nu_\mu, \nu_\tau$) are linear combinations of mass eigenstates ($\nu_1, \nu_2, \nu_3$) and their masses ($m_1, m_2, m_3$) are different from each other, which means that at least two of the masses are non-zero.

The relation between the neutrino flavour and mass eigenstates is described by the Pontecorvo-Maki-Nakagawa-Sakata (PMNS) unitary mixing matrix $U$:
\begin{equation}
\left( \begin{array}{c} \nu_e \\ \nu_\mu \\ \nu_\tau \end{array} \right) = U \left( \begin{array}{c} \nu_1 \\ \nu_2 \\ \nu_3 \end{array} \right)
\end{equation}
The $U$ matrix can be represented using three mixing angles ($\theta_{12},\theta_{13},\theta_{23}$) and the imaginary phase responsible for CP symmetry violation $\delta_{CP}$:
\begin{equation}
	U = \left( \begin{array}{ccc} 1 & 0 & 0 \\ 0 & c_{23} & s_{23} \\ 0 & -s_{23} & c_{23} \end{array} \right) \left( \begin{array}{ccc} c_{13} & 0 & s_{13}e^{-i\delta} \\ 0 & 1 & 0 \\ -s_{13}e^{i\delta} & 0 & c_{13} \end{array} \right)
	\left( \begin{array}{ccc} c_{12} & s_{12} & 0 \\ -s_{12} & c_{12} & 0 \\ 0 & 0 & 1 \end{array} \right)
\end{equation}
where $c_{ij}=\cos\theta_{ij},s_{ij}=\sin\theta_{ij}$ and $\delta=\delta_{CP}$.
The last two independent parameters describing the neutrino oscillations are the neutrino squared mass differences: $\Delta m^{2}_{21}=m^2_2-m^2_1$ and  $\Delta m^{2}_{32}=m^2_3-m^2_2$.

The current values of the known oscillation parameters are as follows\cite{pdg}:
\begin{eqnarray}
\sin^{2} 2\theta_{12}	&	=	&	0.857 \pm 0.024 \nonumber \\
\sin^{2} 2\theta_{23}	&	=	&	> 0.95 \nonumber \\
\sin^{2} 2\theta_{13}	&	=	&	0.095 \pm 0.010 \label{eq:oscParamValue}\\
\Delta m^{2}_{21}	&	=	&	(7.50 \pm 0.20) \times 10^{-5} \ eV^2 \nonumber \\
|\Delta m^{2}_{32}|	&	=	&	(2.32^{+0.12}_{-0.08})  \times 10^{-3} \ eV^2 \nonumber
\end{eqnarray}
The outstanding questions in neutrino oscillations that have not yet been solved are:
\begin{itemize}
	\item the value of $\delta_{CP}$,
	\item the sign of $\Delta m^{2}_{32}$ which defines the neutrino mass hierarchy: $m_1<m_2<m_3$ for normal hierarchy (NH) or $m_3<m_1<m_2$  for inverted hierarchy (IH),
	\item the octant of the $\theta_{23}$ angle ($\theta_{23}<45^\circ$ or $\theta_{23}>45^\circ$).
\end{itemize}

Neutrino oscillations provide evidence for non-zero neutrino masses, something which is often not accounted for in the Standard Model.
This discovery was honoured by the last year's Nobel Prize in Physics and this year's Breakthrough Prize in Fundamental Physics.


\section{The T2K experiment}

\begin{figure}[ht]
\centering
\includegraphics[width=0.8\textwidth]{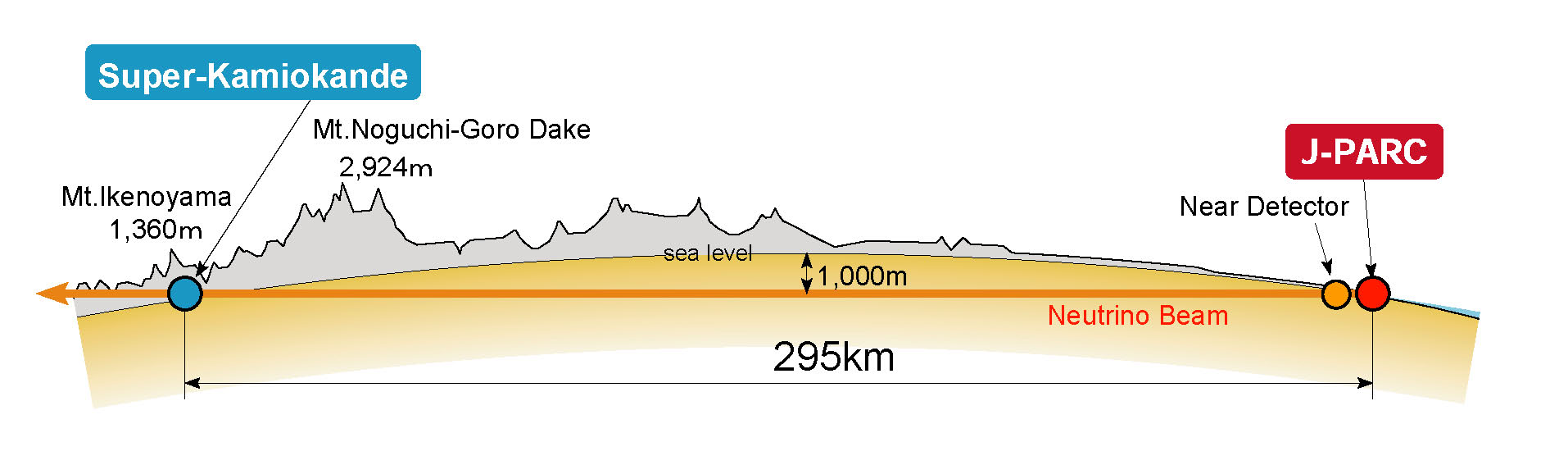}
\caption{Overview of the T2K experiment.}
\label{fig:t2k}
\end{figure}

T2K (Tokai to Kamioka) is a long baseline neutrino oscillation experiment, performed in Japan by an international collaboration of around 500 people from 59 institutions in 11 countries\cite{t2k}. An overview of the T2K experiment is shown in Figure~\ref{fig:t2k}. The muon neutrino and the muon antineutrino beams are produced in the Japan Proton Accelerator Research Complex (J-PARC) in the Tokai village on the east coast of Japan. After 280~m the beam of unoscillated neutrinos goes through the set of two near detectors: INGRID and ND280. After another 295~km, the distance corresponding to the first oscillation maximum, the beam reaches the far detector Super-Kamiokande (SK). The INGRID near detector is placed on the beam axis, while the ND280 near detector and the SK detector are located $2.5^\circ$ away from the neutrino beam axis. The off-axis beam has a narrower energy spectrum and the peak energy to be selected in order to maximise the oscillation probability.

The main goals of the T2K experiment are the measurements of the probability of muon neutrino disappearance, electron neutrino appearance, the investigation of $CP$ symmetry violation/conservation in the neutrino sector and neutrino-nucleus cross-section measurements at T2K beam energies.


\subsection{Neutrino beam}

The T2K neutrino beam is produced in the J-PARC accelerator complex, in which protons are accelerated to 30~GeV and impinge on a graphite target, producing mainly pions ($\sim 90\%$) and kaons ($\sim 10\%$). Then, a set of three magnetic horns focuses positive particles in order to obtain the $\nu_\mu$ beam (neutrino mode) or negative particles for $\bar\nu_\mu$ (antineutrino mode). The positive (negative) pions directed to the decay tunnel produce antimuons (muons) and muon neutrinos (muon antineutrinos). The muons and other particles are stopped by the beam dump and in the ground, and the unimpeded neutrino beam travels farther towards the near detectors and the Super-Kamiokande far detector.


\subsection{Near detectors}

The set of two near detectors (Fig.~\ref{fig:detectors}) is located at the J-PARC site 280~m from the graphite target.

The INGRID detector is located at the centre of the neutrino beam and is composed of iron and plastic scintillator layers.
Its main goal is the day-to-day monitoring of beam direction and intensity. It also serves to measure the neutrino cross-sections at the energies higher than that of the off-axis beam.

The ND280 detector, positioned $2.5^\circ$ away from the beam axis, is a magnetised detector consisting of a number of specialised subdetectors, i.e. Time Projection Chambers and subdetectors containing plastic scintillator layers sandwiched with iron, lead, water layers or other material. ND280 is adjusted to measure the neutrino cross-sections on different nuclear targets, as well as to determine the off-axis neutrino flux and flavour composition at the $2.5^\circ$ angle. These measurements significantly reduce systematic errors in the oscillation analysis.


\subsection{Far detector}

\begin{figure}[ht]
\centering
\includegraphics[width=0.28\textwidth]{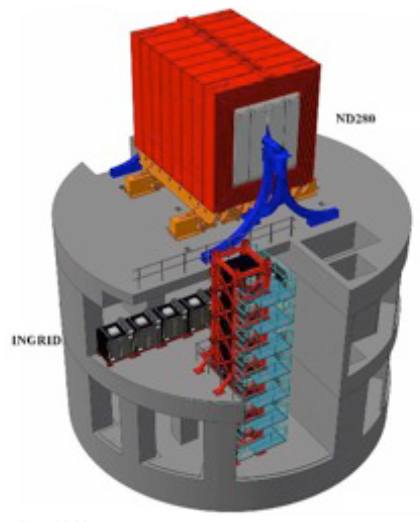}\hspace{2.5cm}
\includegraphics[width=0.23\textwidth]{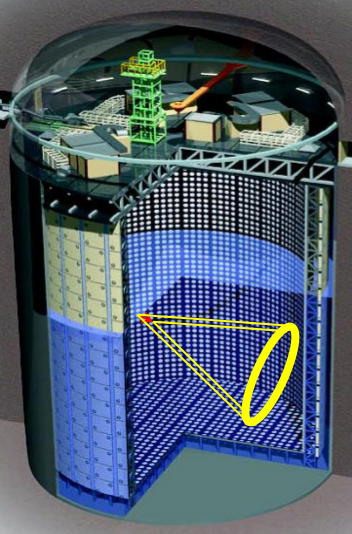}
\caption{Set of the T2K near detectors (left) and the T2K SK far detector (right).}
\label{fig:detectors}
\end{figure}

The Super-Kamiokande detector\cite{sk} (Fig.~\ref{fig:detectors}) is a 50~kton water Cherenkov detector instrumented with approximately 13000 photomultipliers to register charged particles from the neutrino and antineutrino interactions with water. The particle identification is based on the shape of the Cherenkov light rings: sharp for muons and more fuzzy for electrons. The ring analysis can also provide the information about particle directions and energies. The detector has no magnetic field, therefore it cannot distinguish particles from antiparticles.


\section{Oscillation analysis}

The results presented in this paper are based on the data collected by the T2K experiment up to June 2015 which is  $7.0\times10^{20}$~POT for the neutrino mode data and $4.0\times10^{20}$~POT for antineutrino mode data.

For both samples the probabilities of the muon neutrino disappearance and electron neutrino appearance were measured.
First, the expected number and type of neutrino events in the far detector in the case of no oscillations are predicted. These predictions are based on the neutrino flux simulation and model of neutrino interactions constrained by the measurements of the ND280 detector.
The ND280 constraint results in a reduction in the systematic error for the neutrino beam mode analysis from over 20\% to around 7\%\cite{numoderesults}.
Then, based on the ratio between measured and predicted number of SK events as a function of neutrino energy, the oscillation parameters are fitted.
T2K uses fixed solar oscillation parameters: $\sin^2\theta_{12}=0.306$ and $\Delta m^2_{21}=7.5\times10^{-5}$~eV$^2/$c$^4$\cite{solar,numoderesults}.


\subsection{Neutrino oscillation results}

\begin{figure}[ht]
\centering
\includegraphics[width=0.33\textwidth]{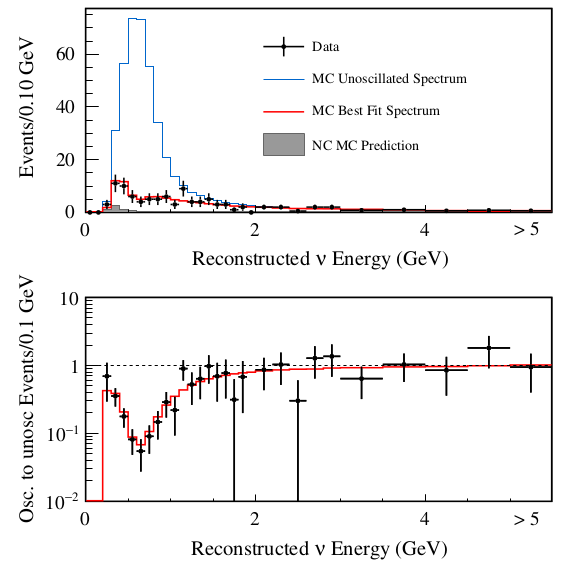}
\includegraphics[width=0.45\textwidth]{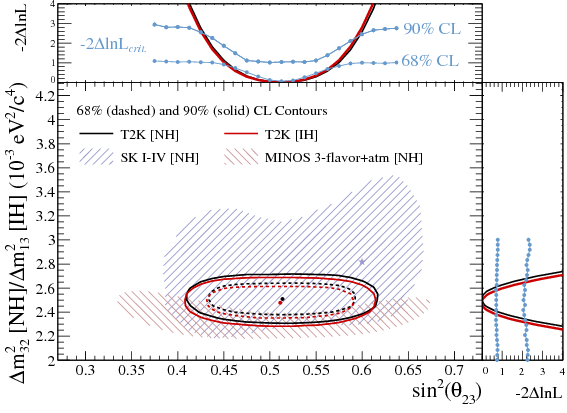}
\caption{The reconstructed neutrino energy spectrum for the data, best-fit prediction, and prediction for no oscillation (top left) and the ratio to the prediction for no oscillation (bottom left)\cite{numoderesults}, as well as the 68\% and 90\% C.L. confidence regions for $\sin^2\theta_{23}$ and $\Delta m^2_{32}$ (NH) or $\Delta m^2_{13}$ (IH) (right)\cite{numoderesults2}.}
\label{fig:numudisapp}
\end{figure}

The difference between the muon neutrino spectra measured and expected without oscillations, is shown in Fig.~\ref{fig:numudisapp}\cite{numoderesults,numoderesults2}. The expected number of the $\nu_\mu$ events without oscillation is equal to $446.0\pm22.5$, while the observed number is 120. T2K has provided the world's leading measurement of the $\theta_{23}$ angle. The best fit values for muon neutrino disappearance are: $\sin^2\theta_{23}=0.514$ and $\Delta m^2_{32}=2.51\times10^{-3}$~eV$^2/$c$^4$ ($\sin^2\theta_{23}=0.511$ and $\Delta m^2_{13}=2.48\times10^{-3}$~eV$^2/$c$^4$) for NH (IH).

\begin{figure}[ht]
\centering
\includegraphics[width=0.33\textwidth]{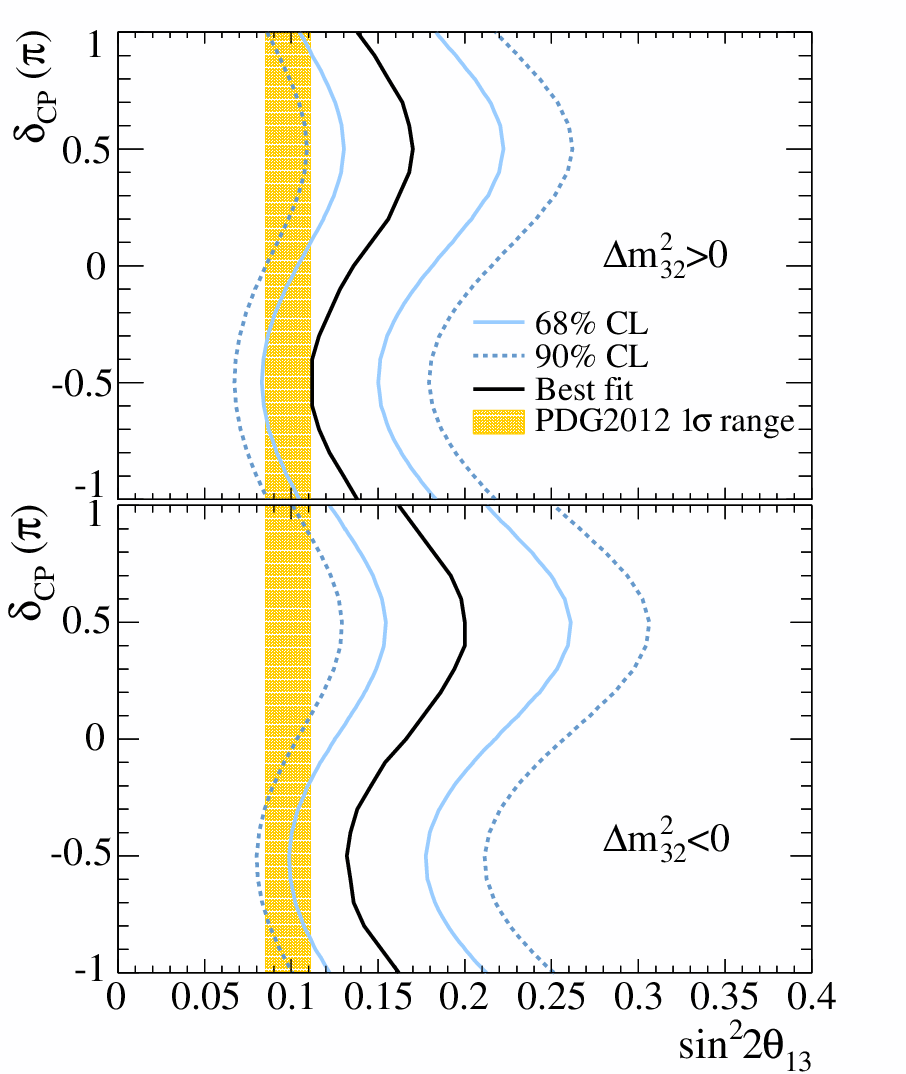}
\includegraphics[width=0.52\textwidth]{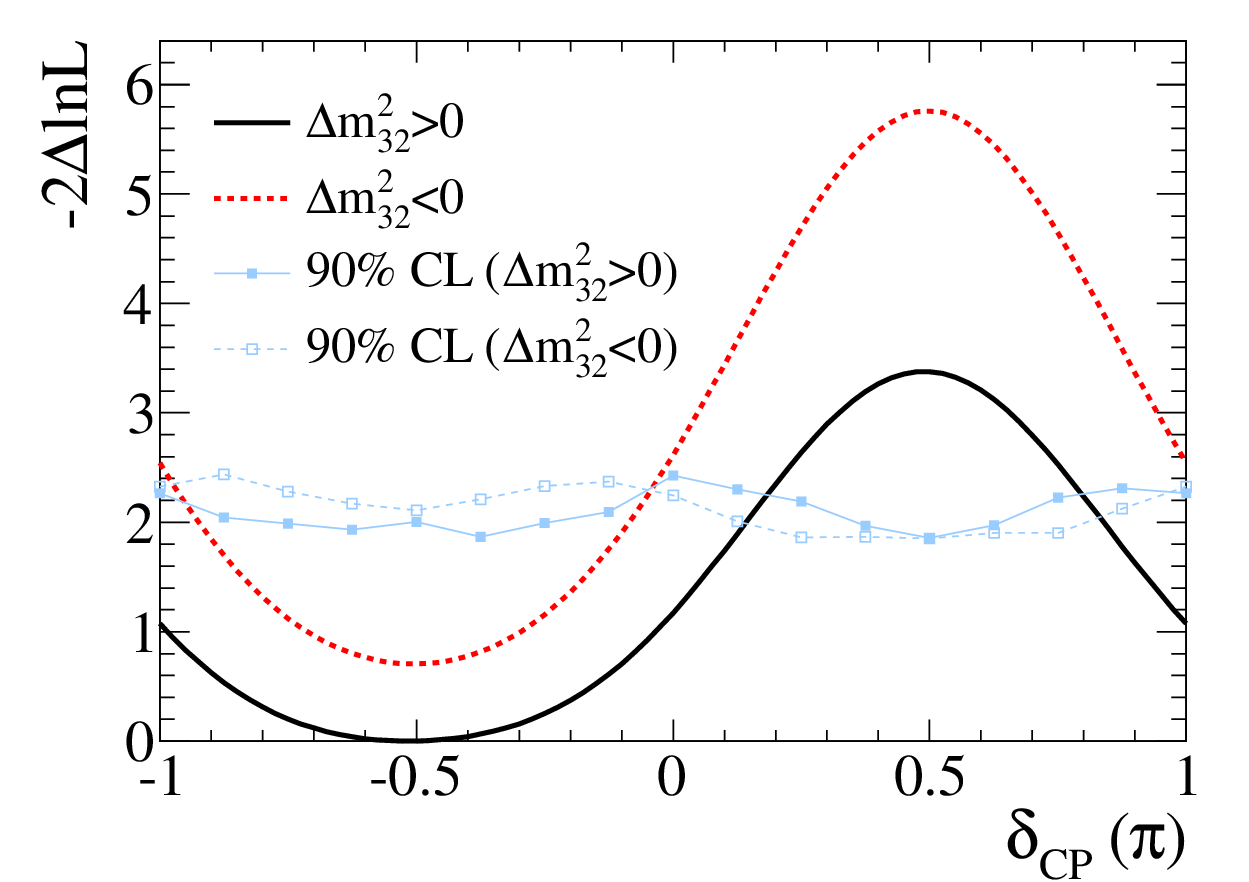}
\caption{The 68\% and 90\% C.L. allowed regions for $\sin^22\theta_{13}$ as a function of $\delta_{CP}$ assuming NH (top left) and IH (bottom left). The $\delta_{CP}$ values above 90\% C.L. (right)\cite{numoderesults3}.}
\label{fig:numeapp}
\end{figure}

For the $\nu_e$ appearance analysis\cite{numoderesults3,numoderesults} $4.92\pm0.55$ events were expected in the case of no oscillation, and the measured number of events was 28. This means that the electron neutrino appearance was confirmed at the level of 7.3$\sigma$. The best fitted value, assuming $\delta_{CP}=0$, is $\sin^22\theta_{13}=0.14$ ($\sin^22\theta_{13}=0.17$) for NH (IH). The allowed regions for $\sin^22\theta_{13}$ as a function of $\delta_{CP}$ are shown in Fig.~\ref{fig:numeapp}. The yellow band shows the average $\theta_{13}$ value from \cite{solar}. Based on the T2K and reactor experiments data\cite{solar,dayabay,reno,doublechooz} the region of $\delta_{CP}$ excluded at 90\% C.L. for NH is $(0.19,0.80)\pi$ and for IH: $(-1.00,-0.97)\pi$ $(-0.04,1.00)\pi$.


\subsection{First antineutrino oscillation results}

\begin{figure}[ht]
\centering
\includegraphics[width=0.33\textwidth]{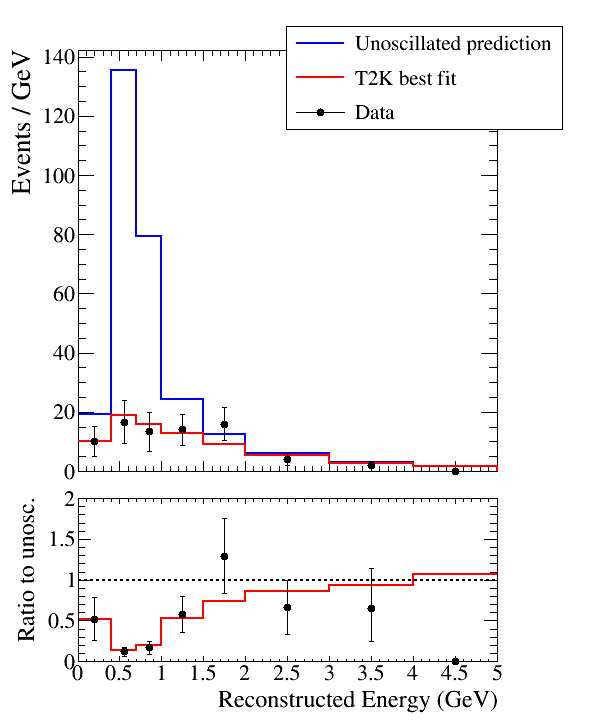}
\includegraphics[width=0.52\textwidth]{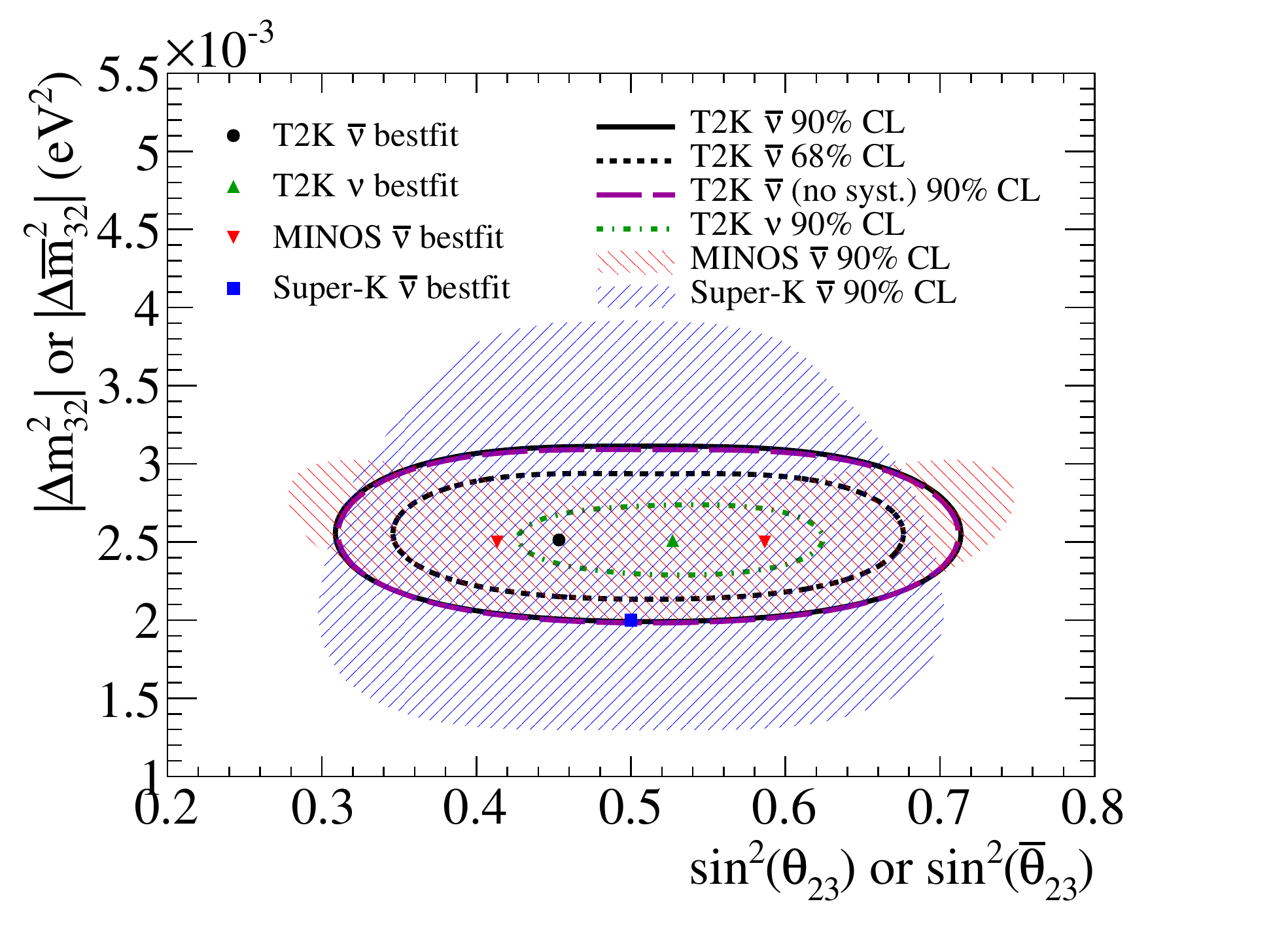}
\caption{The reconstructed antineutrino energy spectrum for the data, best-fit prediction, and prediction for no oscillation (top left) and their ratio to the prediction for no oscillation (bottom left), as well as the 68\% and 90\% C.L. confidence regions for $\sin^2\theta_{23}$ and $\Delta m^2_{32}$ (NH) or $\Delta m^2_{13}$ (IH) (right) for antineutrino beam mode\cite{antinumoderesultsdisap}.}
\label{fig:numubardisapp}
\end{figure}

In antineutrino mode the beam contamination from neutrinos is significantly higher than the antineutrino contamination in the neutrino beam.
As previously stated, the SK detector cannot distinguish particles from antiparticles and, therefore, the selected events contain both neutrinos and antineutrinos ($\nu+\bar\nu$).
The preliminary results for muon antineutrino disappearance in the antineutrino beam mode data were shown in \cite{antinumoderesultsdisap}, the number of predicted $\nu_\mu+\bar\nu_\mu$ events without oscillation is 104 (Fig.~\ref{fig:numubardisapp}). The measured number of such events is 34. The best fit values for the oscillation parameters for antineutrinos are as follows: $\sin^2\theta_{23}=0.515$ and $|\Delta m^2_{32}|=2.33\times10^{-3}$~eV$^2/$c$^4$. With this result, T2K has reached the world's best measurement of $\theta_{23}$ angle in antineutrino mode. The parameter values obtained for neutrinos and antineutrinos are in agreement within the errors, as predicted by $CPT$ conservation.

In the antineutrino mode $\bar\nu_e$ appearance analysis\cite{antinumoderesults}, the predicted number of $\nu_e+\bar\nu_e$ events for the most probable value of $\delta_{CP}=-\pi/2$ is 3.7 (4.2) for normal (inverted) hierarchy. The observed number of events is 3. At this point the statistical uncertainty is too large to draw any conclusions. Collecting more data is crucial here.


\section{Summary and outlook}

The T2K experiment is a leading long-baseline neutrino oscillation experiment that has provided one of the world's best measurements of the $\theta_{23}$ mixing angle for neutrinos and antineutrinos. Both results are consistent with each other as is expected from the $CPT$ conservation.
T2K has observed electron neutrino appearance in a muon neutrino beam at the level of 7.3$\sigma$.
Using $\nu_e$ results combined with the measurements of $\theta_{13}$ from the reactor experiments the first hint of the non-zero value of the $\delta_{CP}$ phase was provided.

To obtain statistically significant results for oscillation parameters in antineutrino mode it is planned to increase the antineutrino beam statistics by an order of magnitude and perform a joint neutrino and antineutrino 3 flavour analyses.


\section*{Acknowledgements}

I thank the T2K collaboration for selecting me to present these results.
This work was partially supported by the Polish National Science Centre, project number: UMO-2014/14/M/ST2/00850.


\end{document}